# Reference levels, signal forms and determination of emission factor in DLTS

Hoang Nam Nhat
*Faculty of Physics, Vietnam National University*
*334 Nguyen Trai, Thanh Xuan, Hanoi, Vietnam*
*E-Mail: namnhat@gmail.com*

**Abstract.** The existence of reference levels of signals which determine directly the temperature dependence of emission factor in deep level transient phenomena is discussed. The basic algebraic structure of reference levels in the classical DLTS is studied and various signal forms with derived reference levels are given. We then demonstrate the use of these signal forms and compare them with the classical DLTS double boxcar signal.

## 1. Introduction

The existence of the deep levels is an important phenomenon in semiconductor physics. It is well-known that they cause many considerable behaviours of materials. The characterization of the deep traps faced many difficulties until 1974 when Lang has introduced a spectroscopic method called the Deep Level Transient Spectroscopy (DLTS) [1]. This allows to deduce from the exponential capacitance decays $C(t) = \Delta C e^{-e_n t}$ the basic physical parameters of the traps such as the activation energy, capture cross-section and concentration. The Lang's method has been widely accepted today as the standard tool, although it has several limitations such as the slow run and relatively low resolution. To extract the trap parameters from the exponential decays, Lang has introduced the *signal form* S(T)=$C(t_1)$–$C(t_2)$ - technically realized using a double boxcar circuit, which monitors the capacitance transients at two different times. This function S(T) has a desirable property that it shows *maximal gain at certain temperature* related to the double boxcar rate windows setting. So by scanning the S(T) over temperature several times one obtains the functional dependence of emission factor on temperature *e=f(T)* and can construct the Arhenius plot ln($e$/$T^2$) versus 1000/T for the determination of trap parameters (Fig.1). The key element in this technique is thus the determination of the temperature dependence *e=f(T).*

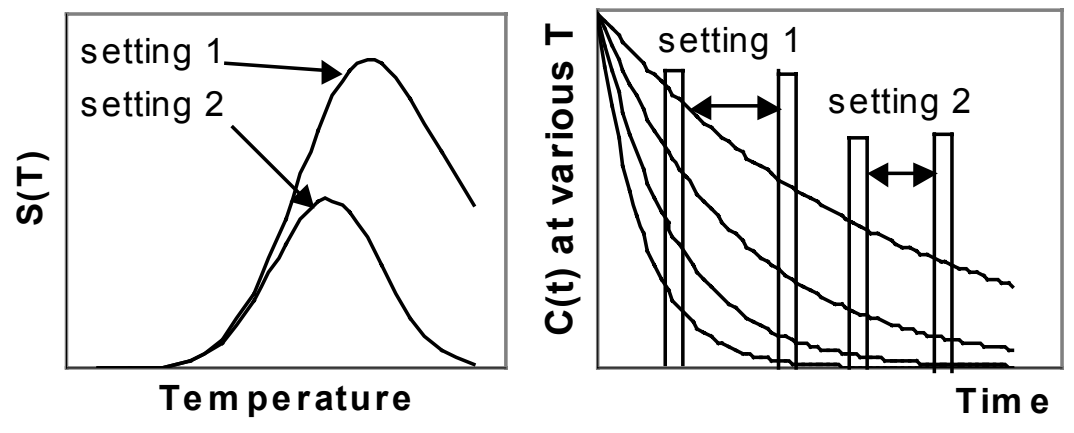

**Fig.1**. Lang's method scans S(T)=C($t_1$)–C($t_2$) for various $t_1$ and $t_2$ settings and draws the temperature dependence of S(T). The maximum determine the temperatures T of the emission factor $e_{max}$ set forth by the rate windows.

Up to now, many attempts have been made in this field to improve the DLTS method. Among the techniques that have been reported [2-14] (the list is certainly not complete), there are two that attracted general attention: *the Fourier and the Laplace technique*. These are both transformation methods manipulating with the whole range of measured data, usually digitally recorded 512 or 1024 points. Recall that the classical S(T) uses only 2 points and throws the rest away. In general the Fourier and the Laplace signal forms show more sensitive peak structure of the gain, but since they do not involve any rate window the exact emission factor at the maximal gain can not be calculated in advance. Thus the correspondence of the peaks and the deep centers appears in these cases somehow subtle and arbitrary.

A common feature of all spectroscopic methods is the presentation of the analytic algorithm converting the set of the capacitance transients *C(t)*, each of them has been recorded at some preset temperature T, into the specific values of certain analytic functions $f_n$(T), showing the peak structures according to T. The $f_n$(T) have two important properties: (1) they are *spectroscopic* in the context that each of the peaks in $f_n$(T) can be associated with one specific deep center and (2) they are *linear,* i.e. the Arhenius plot [ln($e$/$T^2$) versus 1000/T] transformation of the maxima of arbitrarily chosen peak is linear. The functions $f_n$(T) represent the algorithm and usually the method is named after $f_n$(T). Hereinafter the $f_n$(T) are refered to as the *signal form*. For short we may remove the index $n$ denoting the time-settings and use $f$(T) instead of $f_n$(T). The different signal forms involve the different number of measured data and have the different ability in separation of the overlapping deep centers. The classical Lang's signal form, for example, involves only 2 points in the whole transient, whereas the Fourier and the Laplace signal forms are composed principally of the whole transient. There is not known until today any other spectroscopic signal form than the above three.

In this work we present the study of the algebraic structure of the Lang's classical signal form S(T) showing that this form possesses a desirable property of having a so-called *reference level* of signal which directly determines the relationship $e$=$f$(T). This property of DLTS was not reported anywhere before. We then introduce the classes of many other signal forms having the same algebraic structure of the reference levels and reducing the Lang's form as a special case. In contrast to the Lang's form that involves 2 values of $C(t)$, there is a class of forms which involve only 1 single value $C(t)$. This is a surprising fact these forms also provide the peak structure of gain according to T. The Lang's signal form is extended into the class of signal forms which contains many other forms providing the same results as the Lang's form. The fact that there exist many analytic functions $f$(T) fulfilled the requirement of being the signal forms is first described in this paper.

## 2. The reference levels in Lang's signal form S(T) and their algebraic structure

The dependence of the capacitance transient $C(t)$ on time $t$ is considered in general case as:

$$C(t) = C_0 + \sum \Delta C_i e^{-e_i t} \tag{1}$$

where $C_0$ is $C(t{=}\infty)$, $\Delta C{=}\sum\Delta C_i = C(t{=}0){-}C_0$ and $i$ denotes the number of present deep traps.

With respect to the normalized capacitance given as $C_n(t){=}(C(t){-}C_0)/\Delta C$, and denote $t_1{=}t{-}d,\ t_2{=}t{+}d,$ we redefine the Lang's signal for this general case:

$$S(T) = C_n(t-d) - C_n(t+d) = \sum (\Delta C_i / \Delta C)[e^{-e_i(t-d)} - e^{-e_i(t+d)}] \tag{2}$$

Suppose that the traps are independent and not overlapping each other (they are far each from other in the temperature scale), one may differentiate this signal according to some emission factor $e_i$, leaving the other ones zeroed, to determine the signal maximal gain in the given temperature range. We modify the result from [1] with respect to the variables $t$ and $d$ mentioned above:

$$e_{max} = ln[(t+d)/(t-d)]/2d \tag{3}$$

This relation shows that by fixing the rate windows (by $t$ and $d$) one also selects the emission factor to which the Lang's signal reacts mostly when it scans through the set temperature range. With the increase of temperature the trap begins to release electrons and it releases mostly when the emission factor is high enough, raising the Lang's signal to maximum. But when the trap becomes blank, the emission process slows down resulting in the drop of Lang's signal. This intuitive understanding of the emission process - although not correct, offers certain physical meaning to the Lang's signal and set the believe that it really depicts the physical traps.

One thing that seems either unobserved or attracted no considerable attention from the Lang's time is that the relation (3) used to obtain the $e_{max}$ almost equals $1/t$ numerically. Using the Euler number definition formula $\lim_{n\to\infty}(1+1/n)^n = e$ one can without difficulty prove that $ln[(t+d)/(t-d)]/2d$ really converges to $1/t$ when $d \to 0$. Giving the fact that $ln[(t+d)/(t-d)]/2d \sim 1/t$, the $e_{max}$ always corresponds to $C_n(t){=}e^{-1}$ (e is Euler number). This special feature of the classical double boxcar technique is illustrated in Fig.2, where one can see that the $e_{max}$ occurs exactly when $C_n(t)$ passes through the cross-point of the gate central position $t$ and the line $C_n{=}e^{-1}$. This means that despite of the variation in the rate window positions, the only area of importance was $C_n(t){=}e^{-1}$. The evident consequence follows immediately that to detect the functional dependence of the emission factor on the temperature $e_i{=}f(T)$ one simply check the cross-points of $C_n(t)$ and $C_n{=}e^{-1}$ to obtain directly the value of emission factor ($e_i{=}1/t$) corresponding to the given temperature T. For this reason we call $C_n{=}e^{-1}$ the *reference level* of the signal form S(T). It is a great advantage for the signal form to possess the reference level since this means that $e{=}f(T)$ can be derived directly from its reference level.

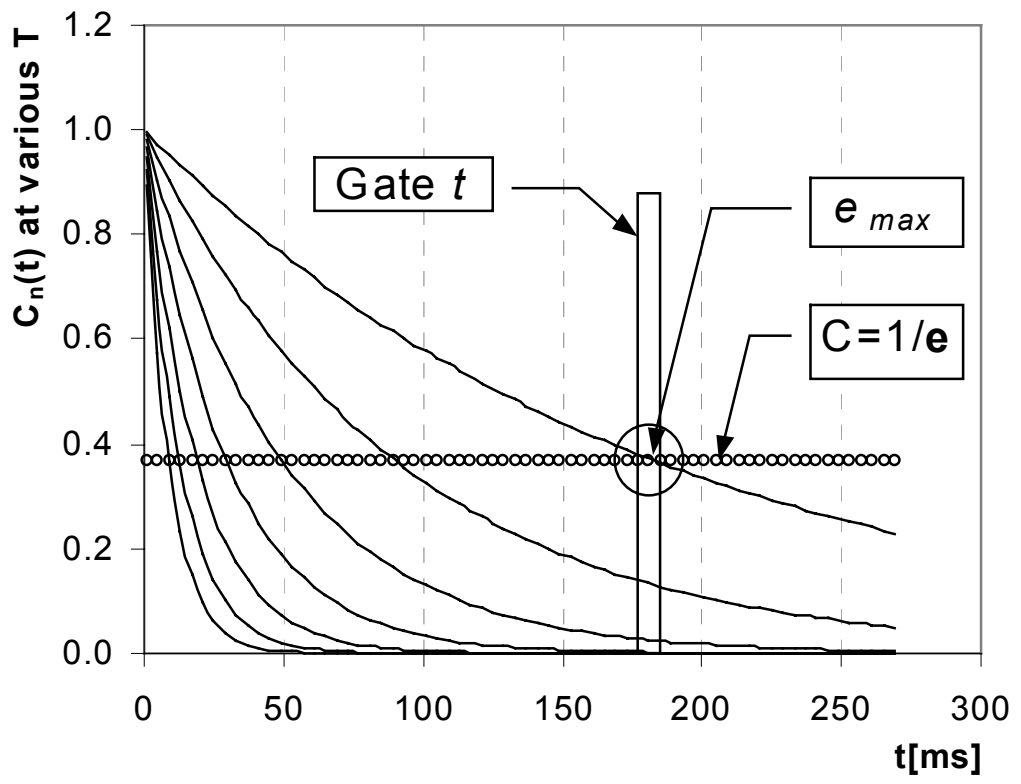

**Fig.2**. The special feature of the double boxcar technique: the rate window [$t{-}d,\ t{+}d$] shows maximum according to T when the $C_n(t)$ decreases through the area $C_n(t){\sim}1/e{=}0.368$.

Although the Lang's signal only approaches this reference level in the limit case when the gate width $2d$ is infinitesimally small, there is a lot of other signal forms as discussed in the next section, which have exact reference level. The importance of reference levels follows from the fact that they lead to the understanding of the algebraic structure of the exponential decays in general and of the capacitance transient particularly. We now introduce the so-called *Lang's signal class* and derive the algebraic structure for this class.

Consider the moving of gate from $t$ to $t'{=}at$, for $a$ is a positive real number. Since $e_{max}$ depends inversely on $t$ it follows that the emission factor $e_i(t)$ detected on the basis of $e_{max}(t)$ changes as: $e_i(t') = e_i(at) = 1/at = (1/a)e_i(t)$. The transient associated with this $e_i(t')$ will have at time $t$ the value equal to the value of the transient associated with $e_i(t)$ at time $t/a$:

$$e^{-e_i(t')t} = e^{-e_i(t)t/a} = C_n(t/a) = [C_n(t)]^{1/a}$$

So we can construct a *modified Lang's signal*, to be called of the *order a* as:

$$S(T)^{[a]} = C_n(t-d)^{1/a} - C_n(t+d)^{1/a} \tag{4}$$

which still has a central position at $t$ but produces the maximal output along the reference level $C_n$=e$^{-a}$ (e=2.718282). Of course, the classical Lang's signal S(T) is of order 1: S(T)[1]. With all possible $a$, the system S(T)[a] forms a class of signals - the *Lang's signal class*. The fact that the $e_{max}$ of S(T)[a] really converts to $a/t$ when $d \rightarrow 0$ can also be observed by differentiating S(T)[a] according to $e_i$ (leaving all other $e_{j\neq i}$ =0) and set it to 0. The result is: $e_{max}$(S(T)[a])= $a\ ln[(t+d)/(t-d)]/2d = ae_{max}$(S(T)[1]) = $a/t$. When $a$<1, the S(T)[a] catchs $C_n$=e$^{-a}$ at lower T and when $a$>1 it catches $C_n$=e$^{-a}$ at higher T compared to S(T).

This signal class associates each point $X$ in the plane [$y=C_n(t)$, $x=t$] with some horizontal reference level line $y=e^{-a}$ and the vertical line $x=t$, so that $X$ lies in the intersect between these two lines. Each point $X$ thus determines a unique emission factor $e_i=a/t$. It is naturally to unify $X$ with $e_i$ and write $e_i=e_i(a,t)$. From the analysis above it is obvious that:

$$e_i(a,t)= ae_i(1,t)= e_i(1,t/a) \qquad (5)$$
$$e_i(a,t)^n= a^n e_i(1,t)^n=a^n e_i(1,t^n)= e_i(a^n,t^n)=e_i(1,(t/a)^n)$$

This tells us about the equivalence of all reference levels in the signal processing system using the double boxcar. The following relations comes straightforward.

$$\lambda[e_i(a,t)+e_i(b,t)] = \lambda e_i(a,t)+\lambda e_i(b,t)= \qquad (6)$$
$$= \lambda a e_i(1,t)+\lambda b e_i(1,t)=\lambda(a+b)e_i(1,t)= e_i(\lambda(a+b),t)$$
$$[e_i(a,t^n)\times e_i(b,t^m)]^\lambda = e_i(a,t^n)^\lambda \times e_i(b,t^m)^\lambda =$$
$$= a^\lambda e_i(1,t)^{n\lambda}\times b^\lambda e_i(1,t)^{m\lambda} =(ab)^\lambda e_i(1,t)^{\lambda(n+m)} =$$
$$= e_i((ab)^\lambda,t^{\lambda(n+m)})$$

One may notice that they follow a linear algebra on $\Re^2$.

## 3. The signal classes and forms

There is an important property of the Lang's signal form: it shows certain separability when the different traps overlap. The signal that is worth the use in practice should be both spectroscopic and resoluble. Up to now, the only spectroscopic signals that brought better resolution were from the transformation of the whole transient. These signals, however, do not possess the reference levels and their algebraic structures are quite different.

This section describes two classes of the signal forms, which we call here the Gaussian and the Poisson class (to the later one the Lang's class S(T)[a] reduces as a special case), possessing the same algebraic structure of the reference levels as the Lang's signal form and also fulfilling the requirement of being resoluble and spectroscopic. The fact that there may exist other spectroscopic signals than the Lang's one can be intuitively recognized from the temperature dependence of $C(t)$ (Fig.3). The simplest way how to creat the peak-shape function from the $C(t)=f$(T) is to either differentiate $C(t)$ according to T (or done by Lang, to substract $C(t_2)$ from $C(t_2)$ - which evidently reduces to the differentiation when the $C(t)$-s become infinitesimally close). These classes are summarized in the Table 1, where the last column shows the estimation for maximal pseudo-random noise level (in % of the maximal signal) that does not disturb their $e_{max}$ more than 5% from the correct value.

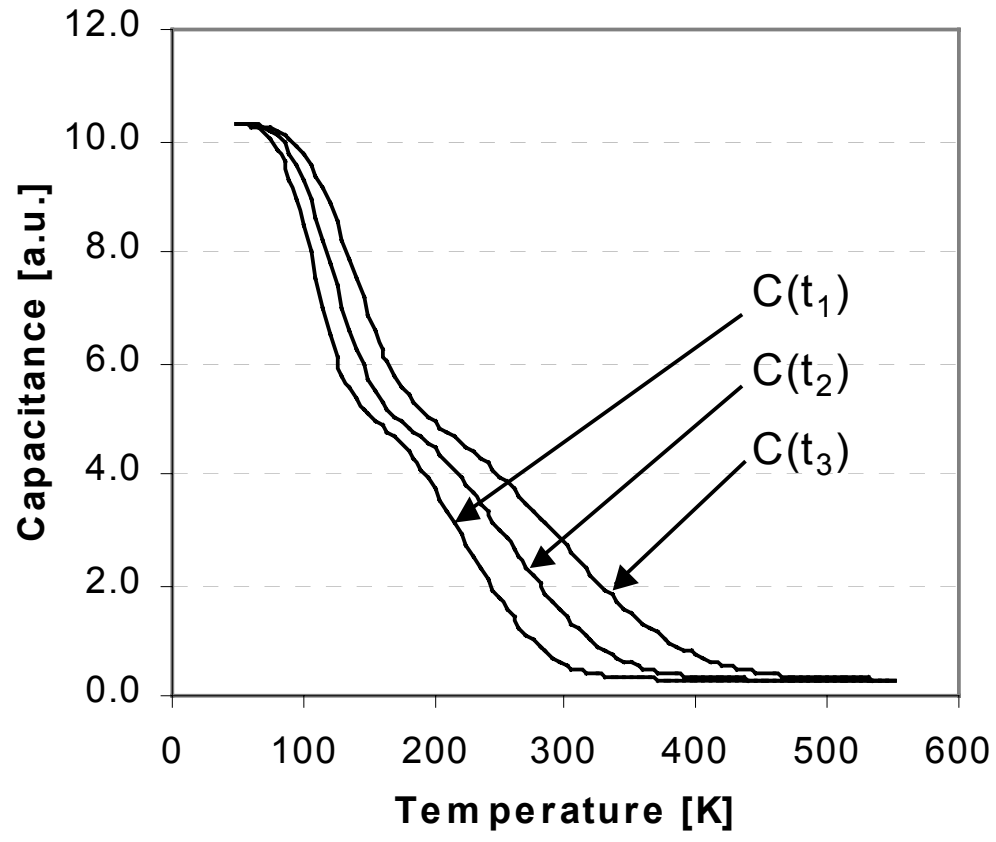

**Fig.3**. The development of capacitance at three successive times for the Lang's *n*-GaAs example with two traps E=0.44eV and 0.75eV.

In general, the signal classes can be classified into two different groups. The 1st is the *finit element group*, consisting of the classes with signals formed from the finit number of $C(t)$. The 2nd is the *infinite element group* consisting of the classes with signals formed from the infinite number of $C(t)$. This classification can be extended to cover also the 3rd class of signal forms, which deal with the non-analytic algorithms, that is the *fractal group*. Principally, any non-analytic algorithm $F(t,T,C(t,T))$ taking $C(t)$, $t$, T as the inputs and outputs the peaks can be considered as the signal form if it satisfies the conditions for the signal forms. The study on the 2nd and 3rd groups will be presented in another paper. This work set focus on the 1st group of signal forms.

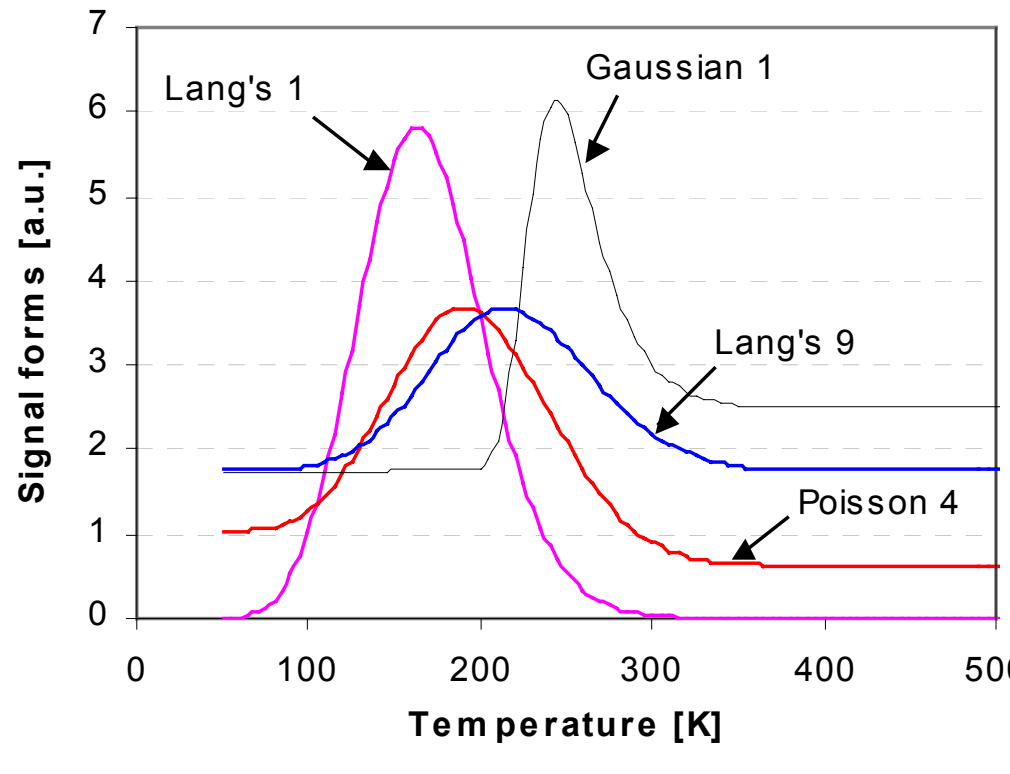

**Fig.4**. Comparison of some selected signal forms to the classical Lang's S(T) form for a sample with one trap E=0.44eV.

***Table 1***. *The finit element signal classes: signal forms, their $e_{max}$ and reference levels*

| *Class* | | *Signal forms* | $e_{max}$ | *Reference level* | *Max noise* |
|---|---|---|---|---|---|
| Gauss (unitary) | 1 | $\beta C(t) - C(t)^{\alpha}$<br>$e^{[\beta C(t)-C(t)^{\alpha}]}$<br>*usually β=5-10* | for $\alpha=2$,<br>$e_{max}=(1/t)ln[2\Delta C/(\beta-2C_0)]$ | $e^{-a}$,<br>$a=ln[2\Delta C/(\beta-2C_0)]$ | 1.5-2% |
| | 2 | $C(t)e^{\beta C(t)-C(t)^{\alpha}}$ | for $\alpha=2$,<br>$e_{max}=(1/t)ln[2\Delta C/(1+\beta-2C_0)]$ | $e^{-a}$,<br>$a=ln[2\Delta C/(1+\beta-2C_0)]$ | 1.5-2% |
| | 3 | $ke^{-(C(t)-\mu)^2/2\sigma^2}$<br>*usually* $\mu\sim1$, $2\sigma^2=0.2$<br>*k only scales the graph* | $e_{max}=(1/t)ln[\Delta C/(\mu-C_0)]$ | $e^{-a}$,<br>$a=ln[\Delta C/(\mu-C_0)]$ | 1.0-1.5% |
| Poisson (unitary) | 4 | $-C(t)\ln[\alpha C(t)]$<br>*for* $0<\alpha<1$, *usually* $\alpha=0.2$ | $e_{max}=(1/t)ln[\alpha\Delta C/(e^{-1}-\alpha C_0)]$ | $e^{-a}$,<br>$a=ln[\alpha\Delta C/(e^{-1}-\alpha C_0)]$ | 3-5% |
| | 5 | $C(t)^{\beta}\lambda^{-\alpha C(t)}$<br>*for* $\lambda>1$, *usually* $\lambda=2$ | $e_{max}=(1/t)ln[\Delta C ln\lambda/(1-C_0 ln\lambda)]$ | $e^{-a}$,<br>$a=ln[\Delta C ln\lambda/(1-C_0 ln\lambda)]$ | 3-5% |
| Lang (binary) | 6 | $C_n(t_1)^{1/a} - C_n(t_2)^{1/a}$<br>*need normalized* $C_n(t)$ *but not for* $a=1$ | $e_{max}=a\ ln(t_1/t_2)/(t_1-t_2)\sim a/t$ | $e^{-a}$ | 1-1.5% |
| | 7 | $C(t_1)^n / C(t_2)^n$<br>*usually n=1 or 2* | for $t_2=2t_1$:<br>$e_{max}=(1/t)\ln(1+\sqrt{1+\Delta C/C_0})$ | $e^{-a}$,<br>$a=\ln(1+\sqrt{1+\Delta C/C_0})$ | 0.5% |
| | 8 | $\alpha\Delta C-(C(t_1)+1/C(t_2))$<br>*can not be used with the normalized* $C_n(t)$ | for $t_2=t_1=t$ *(unitar signal)*:<br>$e_{max}=-(1/t)\ln(-1+\sqrt{1+1/C_0^2})$ | $e^{-a}$<br>$a=-\ln(-1+\sqrt{1+1/C_0^2})$ | 1-1.3% |
| | 9 | $C_n(t_2)\ln C_n(t_1)-C_n(t_1)\ln C_n(t_2)$<br>*need normalized* $C_n(t)$ | estimation for $t_2=2t_1$ :<br>$e_{max}=1.21188215/t$ | $e^{-a}$,<br>$a=1.21188215$ | 1-1.3% |

*The finit element signal classes*

The signal forms are composing from one single $C(t)$ or from a finit number of $C(t_i)$. The Lang's class is a special case where the number of $C(t_i)$ is 2. It is worth to adopt the following notation. According to the number of $C(t_i)$ they consist of the signal form is called the unitary or binary signal form.

Among the unitary signal forms, the Poisson ones - derived from the Poisson distribution function, deserve most attention since they provide sharp peak and their resistibility to noise is high. The Gaussian forms also possess good peak structure but they seem more sensitive to noise. Both these two classes are of $e^{-a}$ reference level class with $e_{max}=a/t$. Fig.4 compares some of them with the classic Lang's form which belongs to the middle quality signals. The Lang's signal form, workable in the interference of 1-1.5% noise, is the best form among the binary ones but is comparable to the Gaussian forms (1.5%) and is worse than the Poisson forms (3-5%).

A common feature of the finit element forms is that they all have $e^{-a}$ reference level with $a$ preset. The $e_{max}$ depends only on $t$ and is always $a/t$. This enables the straightforward construction of the functional dependence $e=f$(T): at each T when the $C(t)$ is recorded, the time $t$ where $C(t)$ crosses the horizontal line C=$e^{-a}$ determines $e$(T)=$a/t$. So the repeated scanning of $C(t)$ over the whole temperature range as for the classical DLTS is not needed. The use of the unitary signal forms even makes the measurement process more faster in one aspect that we don't need to scan the whole time $t$ and can set focus onto the specific area. This topic is however the subject of the further study. The existence of the unitary signal forms itself is a surprising fact. Fig.5 illustrates the use of the Gaussian signal form to determinate the traps in the Lang's example $n$-GaAs.

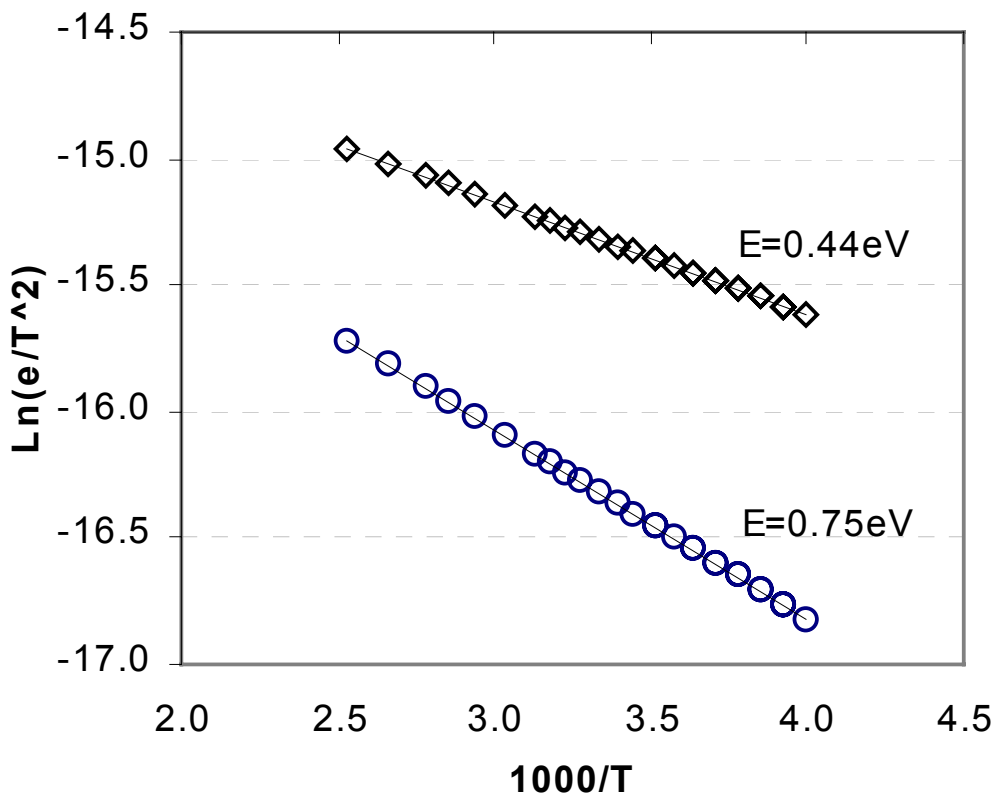

**Fig.5**. The Arhenius plot constructed using the Gaussian signal form No. 1 (Table 1) for the Lang's example $n$-GaAs with two traps E=0.44eV and 0.75eV.

## 4. Conclusion

The existence of reference levels of signals and many signal forms in DLTS is discussed here for the first time. We showed that the set of the reference levels forms a linear algebra which holds valid for the presented classes of signal forms. The reference levels allow the direct determination of $e=f(T)$ in a geometrical way. Besides the Lang's signal class, obtaining from the modification of the Lang's classical form S(T), the two other signal classes - the Gaussian and the Poisson classes, are discussed. The existence of a unitary class of signals is probably the most interesting result of this work. The unitary signal forms are, in one hand, more persistent to noise, in the other, reduce the need of repeating the measurement. They provide very good results compared to the classical DLTS.